\documentclass[pre,aps,superscriptaddress,showpacs]{revtex4}
\usepackage{amsbsy}
\usepackage{amssymb}
\usepackage[dvips]{graphicx}

\begin{document}
\title{ The Statistics of the Number of Minima in a Random Energy Landscape}

\author {Satya N. Majumdar}
\affiliation{
CNRS; Univ. Paris Sud, UMR8626, LPTMS, ORSAY CEDEX, F-91405, France}

\author {Olivier C. Martin}
\affiliation{
CNRS; Univ. Paris Sud, UMR8626, LPTMS, ORSAY CEDEX, F-91405, France}

\date{\today}

\begin{abstract}
We consider random energy landscapes constructed from 
$d$-dimensional lattices or trees. The distribution of the number
of local minima in such landscapes follows a large deviation principle
and we derive the associated law exactly for dimension 1. Also of 
interest is the probability
of the maximum possible number of minima; this probability
scales exponentially with the number of sites. We calculate
analytically the corresponding exponent for the Cayley tree
and the two-leg ladder; for 2 to 5 dimensional hypercubic lattices,
we compute the exponent numerically and compare to the 
Cayley tree case.
\end{abstract}
\pacs{02.50.Cw (Probability theory), 05.50.+q}
\maketitle

For many materials that are glassy~\cite{BarratFeigelman02}
local minima 
of the energy (or of the free energy) trap the system for 
long times, leading to
subtle equilibrium and out of equilibrium properties. 
Energy landscapes provide a simple conceptual framework
for modeling these systems but in fact
their use goes much beyond that. For instance the vacua of string
theories are expected to proliferate enormously,
and the problem of estimating the number of local 
minima~\cite{Susskind03, AazamiEasther06}
of string energy landscapes is still an open problem. 
Other examples include rugged (random) energy landscapes
in evolutionary biology~\cite{Gavrilets04}, 
quantum cosmology~\cite{Mersini-Houghton05}, 
manifolds in random media~\cite{Halpin-HealyZhang86},
glassy systems~\cite{MezardParisi87b,BarratFeigelman02}
random potentials~\cite{CavagnaGarrahan99,Fyodorov04} 
and the associated problems in random matrix 
theory~\cite{CavagnaGarrahan00,AazamiEasther06,DeanMajumdar06}.
Typically these systems consider a particle, a configuration of
particles, or even a manifold, subject to a random potential. One extreme 
case for its visual simplicity is that of a point particle in 
a random Gaussian potential; on the opposite extreme are
energy landscapes associated with the configuration of a many
body system. Examples in this last category include:
(1) the $p$-spin glass model~\cite{GrossMezard84}, which in the limit
of large $p$ reduces to the far simpler random energy
model of Derrida~\cite{Derrida80}; (2) atomic
clusters~\cite{Wales99} and other glassy systems with
no quenched disorder~\cite{Angell95}; (3) random manifolds in
random media~\cite{Halpin-HealyZhang86}.

For any landscape, it is desirable to know the statistical properties
of the minima (or more generally of the saddle points).
A quantity frequently considered is the expected number
of minima as a function of their energy~\cite{CavagnaGarrahan99}.
It may also be of interest to consider how the set
of minima are organized topologically, e.g., whether the
barrier tree~\cite{FontanariStadler02} is ultrametric. Our focus
here is to better understand the statistics of
the total number of local minima in 
random energy landscapes; our underlying space
is a regular (Euclidean) lattice on each site
of which resides a random energy.
Let $M$ denote the total number of local minima for given 
values of the energies on each site. Evidently $M$ is a random 
variable as it varies
from one realization of the landscape to another. We are
interested in the statistics
of $M$ as a function of the number $N$ of lattice sites. Similar 
questions were studied recently
in the context of random 
permutations~\cite{Stembridge97,OshaninVoituriez04},
ballistic deposition~\cite{HivertNechaev05},
and in simple models
of glasses~\cite{BurdaKrzywicki06}.
In this paper, we provide a number of analytical results on the distribution
of the total number of local minima for random energy 
landscapes on several lattices; the moments of $M$ are 
easily derived, so our focus concerns mainly
the probability of \emph{large deviations} of $M$ from its mean value, i.e., the
probabilities of atypical configurations. Furthermore we ask what is the
probability of $M$ being at its maximum, corresponding to the
limit where the minima are maximally packed on the lattice.

The paper is organized as follows. In Sect.~\ref{sec:model}
we specify the model and show that the statistical
properties of the minima are independent of the 
individual distribution of energies as long as these
energies are independent from site to site and are drawn from
a continuous distribution.
In Sect.~\ref{sec:GeneralProperties} we cover some
of the simplest properties of the statistics of $M$ and
formulate the large deviation principle.
Then we derive the closed form expression for
the large deviation function in the case of the one dimensional
lattice in Sect.~\ref{sec:OneDimension}. The focus of the
rest of the paper is the maximum
packing problem. In Sect.~\ref{sec:MaxPack}
we determine the probability of maximum packings 
for several solvable cases, namely Cayley trees
and a two-leg ladder. The case of
$d$-dimensional hypercubic lattices is then treated
by numerical computation in Sect.~\ref{sec:Numerics}.
Finally, these different results are discussed 
in Sect.~\ref{sec:Conclusions} and some closing
remarks are given.

\section{The model}
\label{sec:model}

We start with a regular lattice; on each site
$i$ lives a random energy $E_i$, drawn independently from site to site from 
a common distribution $\rho(E)$. 
We assume $\rho(E)$ is continuous and normalized to
unity, i.e. $\int_{-\infty}^{\infty} \rho(E) dE=1$. 
For any choice of the set of $E_i$'s we obtain
an energy landscape, i.e., a topological space
(defined through nearest neighbors on the lattice) with an energy for
each element.
For a given realization of the landscape, a site $i$ is a local minimum
if $E_i<E_j$ where $j$ denotes any nearest neighbour sites of $i$.
Hereafter we shall denote by $M$ the number of local minima on
this landscape.

For any given realization of
the landscape, one can formally express $M$ as
\begin{equation}
M = \sum_{i=1}^N \, \prod_{j/<ij>} \theta\left(E_j-E_i\right)
\label{M1}
\end{equation}
where the product runs over all nearest neighbours $j$ 
of site $i$ and $\theta(x)$
is the Heaviside theta function. We are interested in computing the probability
distribution $P(M,N)$ of the total number of local minima (or equivalently
that of the maxima).

The first important observation is that the distribution $P(M,N)$ is completely
independent of the energy distribution $\rho(E)$. To see this, we formally
express the distribution as
\begin{equation}
P(M,N)= \int\,\ldots \int\, \delta\left(M-\sum_{i=1}^N \, 
\prod_{j} \theta\left(E_j-E_i\right)\right)\, 
\prod_{k=1}^N 
\rho(E_k)\, dE_k
\label{M2}
\end{equation}
where $\delta(x)$ is the Dirac delta function. Next we 
make a change of variable for each $k$
\begin{equation}
x_k = \int_{-\infty}^{E_k} \rho(E)\,dE
\label{M3}
\end{equation}
Clearly $x_k$ is a monotonically increasing function of $E_k$. Therefore, 
$\theta\left(E_j-E_i\right)=\theta(x_j-x_i)$. Moreover, since $\rho(E)$
is normalized to unity, the variable $x$ varies from $0$ to $1$ and so 
Eq.~(\ref{M2})
simply becomes
\begin{equation}
P(M,N)=\int_0^1 dx_1 \ldots \int_0^1 dx_N\, \delta\left(M-\sum_{i=1}^N \, \prod_{j} 
\theta\left(x_j-x_i\right)\right). 
\label{M4}
\end{equation}
We see that the energy distribution $\rho(E)$ simply drops out and 
$P(M,N)$, for arbitrary $\rho(E)$, is 
universal and is the same as when the `new' energy 
variable $x_i$ is drawn independently
from a uniform distribution over $x \in [0,1]$. Thus, the model is 
simplified: at each lattice site
lives a random number $x_i\in [0,1]$ drawn independently from a 
uniform distribution. 
We will refer to this model as the random minima model, for which 
we want to compute the distribution $P(M,N)$ of the 
total number of local minima.

Following the mapping to the uniform distribution, it 
follows that the random minima model
is just the continuous version of the permutation generated landscape recently
studied by Hivert et. al.~\cite{HivertNechaev05}. They considered a 
set of integers $\{1,2,3,\ldots, N\}$.
Each of the $N!$ permutations of this set defines a random energy landscape and
they occur with equal probability. Thus the energy at a site 
is now a discrete integer drawn uniformly 
from the set $[1,2,3, \ldots, N]$ with equal
probability $1/N$. Thus the statistics of the minima in the permutation generated
landscape will be identical to that of the random minima model.

\section{General properties of the distribution of $M$}
\label{sec:GeneralProperties}

The mean and the variance of $M$ are relatively
straightforward to compute on an arbitrary lattice since they depend
only on the local properties of the landscape. To see this, let us define
the variable
\begin{equation}
\eta_i = \prod_{j} \theta(x_j-x_i)
\label{mean0}
\end{equation}
where $j$ runs over the nearest neighbours of $i$ 
and the $x_i$'s are independent random numbers
in $[0,1]$ drawn from the uniform distribution. This 
$\eta_i$ is an indicator function which is 1 if site
$i$ is a local minimum and 0 otherwise. Then it follows that
\begin{equation}
M= \sum_{i=1}^{N} \eta_i.
\label{mean1}
\end{equation}
Taking the average
in Eq. (\ref{mean1}) and using the translational invariance of 
the lattice, it follows that
\begin{equation}
\frac{\langle M\rangle}{N}= \big \langle \prod_{j/<ij>} \theta(x_j-x_i)\big \rangle=\int_0^{1}dx_i \left[\int_{x_i}^1 
dx_j\right]^n = \frac{1}{n+1}
\label{mean2}
\end{equation}
where $n$ is the number of nearest neighbours of any site, i.e., 
the co-ordination number of the
lattice. This result also follows trivially from a 
combinatorial argument on the permutation
landscape: consider the site $i$ with its $n$ neighbours. The number 
of ways one can
arrange the integers on these $(n+1)$ sites with the restriction 
that $i$ is a local minimum is
$n!$ clearly. On the other hand, the total number of 
unrestricted configurations is $(n+1)!$ so the
probability that the site $i$ is a minimum is simply $n!/(n+1)! = 1/(n+1)$. 

The calculation of the variance takes a few more steps. Squaring 
Eq. (\ref{mean1}) and taking the average gives
\begin{equation}
\langle M^2\rangle =\sum_{i,j} \langle \eta_i \eta_j \rangle.
\label{var1}
\end{equation}
Note that the $\eta_i$'s are correlated random variables but 
only over a short range. The
calculation of the correlation function $\langle \eta_i \eta_j \rangle $ 
can therefore
be performed by hand (it only involves the calculation of simple 
integrals involving
a maximum of $2n$ sites). For example, in $1$-d, one gets
\begin{equation}
\sigma^2 = \langle M^2\rangle - {\langle M\rangle}^2 
\to \frac{2}{45} N\quad \quad {\rm as}\quad N\to \infty. 
\label{var2}
\end{equation}
This result has been obtained in various contexts before, such 
as in the calculation
of the number of metastable states in $1$-d Ising 
spin glass~\cite{DerridaGardner86} and also in the
context of the occupation time of a non-Markovian sequence 
in $1$-d~\cite{Majumdar02}. Recently,
the variance for the $2-d$ square lattice was computed in the permutation 
model by Hivert et. al.~\cite{HivertNechaev05}. For large $N$, one gets
\begin{equation}
\sigma^2 = \frac{13}{225} N .
\label{var3}
\end{equation}
In general, on any arbitrary lattice, for large $N$
\begin{eqnarray}
\langle M \rangle &= & a N \nonumber \\
\langle M^2\rangle - {\langle M\rangle}^2 &=& b N
\label{mv}
\end{eqnarray}
where $a$ and $b$ are lattice dependent numbers that can be
computed either using integrals or by combinatorics in the
permutation model.

Near the mean $\langle M\rangle$ and within a 
region $|M-\langle M \rangle | =O(\sqrt{N})$,
the distribution $P(M,N)$ is expected to be a 
Gaussian~\cite{HivertNechaev05} with mean and variance given 
in Eq. (\ref{mv})
\begin{equation} 
P(M,N) \sim \frac{1}{\sqrt{2\pi b N}}\, \exp\left[-(M-aN)^2/{2bN}\right].
\label{gauss1}
\end{equation}
However, for $M$ far from the mean, one would expect 
deviations in $P(M,N)$ from the Gaussian
form. In this paper, we are interested in the 
probabilities of such large deviations, i.e.,
we are interested in computing the probabilities of occurrences 
of configurations that are far from typical. On general grounds, 
in the limit $M\to \infty$, $N\to \infty$ but with their 
ratio $M/N$ fixed but arbitrary,
one expects that $P(M,N)$ has the form
\begin{equation}
P(M,N) \sim \exp\left[-N\, \Phi\left(\frac{M}{N}\right)\right]
\label{ld1}
\end{equation}
where $\Phi(y)$ is a large deviation function. Now, on a given lattice, $M$ can
take values from $0$ to a maximal number $M_{\rm max}<N$. The upper limit follows
from the fact that once a site is a local minimum, none of its nearest neighbours
can be a local minimum. Thus there is a nearest neighbour exclusion principle
for the minima. This constraint indicates that one can 
not pack arbitrarily large number
of minima on the lattice and there is an upper bound 
on the number of minima. For example,
on a bipartite lattice, one can pack at most $M_{\rm max}= N/2$ local 
minima, one at
every alternate site. In particular, on a square lattice the minima will be placed
on a checkerboard pattern. Thus, in this case, the large deviation 
function $\Phi(y)$ 
is defined for $0\le y\le y_{\rm max}=1/2$. On the other hand, on a Cayley tree
with $\mu$ number of branches and $N$ sites, we will see 
later that $M_{\rm max}=    
\mu N/(\mu+1)$, thus $y_{\rm max}= \mu/(\mu+1)$. 

As mentioned above, in the vicinity of its mean, i.e., 
for $|M-\langle M\rangle| = O(\sqrt N)$, 
the distribution $P(M,N)$ is Gaussian as in Eq. (\ref{gauss1}).
This indicates that the large deviation function $\Phi(y)$ is quadratic
near $y=a$
\begin{equation}
\Phi(y) \approx \frac{(y-a)^2}{2b}
\label{qua}
\end{equation}
such that $P(M,N) \approx \exp\left[-N (y-a)^2/{2b}\right]
\sim \exp\left[-(M-aN)^2/{2bN}\right]$
has the required Gaussian form. However,
far from the mean, $P(M,N)$ will have non-Gaussian tails, indicating 
a departure of $\Phi(y)$ from the simple quadratic form. 
A knowledge of the function $\Phi(y)$ would
then allow one to compute the probabilities of occurrences of 
atypical configurations, such
as the probability of a configuration with the lowest number 
of minima (e.g. $M=1$) or 
the ones with the maximal 
number of minima ($M=M_{\rm max})$.

A particular focus of this paper will be to compute the 
probability of a maximally
packed configuration, i.e., when $M=M_{\rm max}$. It follows 
from the general form
in Eq. (\ref{ld1}) that the probability of this 
maximal packing should decay exponentially
with the system size for large $N$,
\begin{equation}
P(M_{\rm max}, N) \sim \exp\left[-N \Phi(y_{\rm max})\right] \sim \gamma^{-N}
\label{ld2}
\end{equation}
where $\gamma= \exp\left[\Phi(y_{\rm max})\right]$ is a 
lattice dependent constant. Of course, this expression also
gives the scaling of the fraction of permutations which have
$M_{\rm max}$ local minima in the permutation landscape.
It is then natural to interpret $\ln \gamma$ as the entropy cost per site
to achieve maximally packed minima.

We shall compute the constant $\gamma$ exactly for a number of 
lattices. For the $1-d$ chain, we will show that 
\begin{equation}
\gamma=\frac{\pi}{2}=1.57079\dots
\label{g1d}
\end{equation}
Note that this result appeared before in the context 
of metastable states in $1-d$ Ising spin 
glasses at zero temperature~\cite{DerridaGardner86}. Furthermore, the 
same $(\pi/2)^{-N}$ decay also appeared as the persistence
probability of a $1-d$ non-Markovian sequence~\cite{MajumdarDhar01,Majumdar02}.
For the $1-d$ case, one can also compute the
full large deviation function $\Phi(y)$ (see 
Section~\ref{sec:OneDimension}).  Another 
lattice where we can calculate $\gamma$ 
exactly is the Cayley tree with 
$\mu$ branches where we show that
\begin{equation}
\gamma= \frac{1}{\mu+1}\, B\left(\frac{1}{\mu+1},\frac{1}{\mu+1}\right)
\label{gct}
\end{equation}
where $B(x,y)$ is the Beta function. As 
expected, for $\mu=1$, it reduces to the $1$-d result
in Eq.~(\ref{g1d}). As $\mu$ increases, $\gamma$ increases slowly
and as $\mu\to 
\infty$, $\gamma\to 2$. A Cayley tree with an
infinite number of branches corresponds to a hypercubic lattice in $d$-dimensions in
the $d\to \infty$ limit. Thus, based on these two 
limiting results, one expects the constant $\gamma$ on 
any $d$-dimensional lattice to satisfy the bounds
\begin{equation}
\pi/2 \le \gamma \le 2 .
\label{bounds}
\end{equation}
Our numerical simulations for $2 \le d \le 5$ are consistent with these bounds.
We have also been able to compute $\gamma$ exactly for a two-leg ladder where we show
that
\begin{equation}
\gamma= \left[\frac{9 \alpha^2}{8}\right]^{1/3}=1.57657\dots
\label{glad}
\end{equation}
where $\alpha=1.86635\dots$ is the smallest positive root of the Bessel
function $J_{-1/3}(z)=0$. The result in Eq. (\ref{glad}) is, of course, consistent with
the general bounds in Eq. (\ref{bounds}). 

We now consider these successive cases in detail.

\section{Exact Large Deviation Function in One Dimension} 
\label{sec:OneDimension}

In a $1$-d chain of size $N$, the distribution $P(M,N)$ of the total number of local minima 
was computed exactly by Derrida and Gardner~\cite{DerridaGardner86},
 although they did not calculate
the large deviation function $\Phi(y)$ explicitly. However, from their result for the
generating function, it is easy to derive the large deviation function by a Legendre
transform. For the sake of completeness, we provide here a brief derivation of the $1$-d 
result, albeit by a slightly different method.

It is useful to define the generating function or the partition function 
$Z(z,N)= \sum_M P(M,N) z^M $ where
$z$ is the fugacity or the weight associated with each minimum. 
For simplicity we consider an open chain of size $N$ and 
let $x_N=x$ be
the value of the random variable at the $N$-th site. 
To write a recursion relation for the partition function, it is convenient
to define two restricted partition functions: $Z_1(x,z,N)$ and $Z_0(x,z,N)$
denoting respectively the partition functions conditioned on the fact that
the $N$-th site has value $x_N=x$ and that it is respectively
a local minimum (i.e., $x_N<x_{N-1}$) or a local maximum ($x_N>x_{N-1}$).
Note that since we are considering an open chain, the last site ($N$) has only
one neighbour to its left, namely the $(N-1)$-th site. 
Knowing the restricted partition functions, one can compute the
full partition function from the relation
\begin{equation}
Z(z,N)= \int_0^1 dx \left[Z_1(x,z,N)+ Z_0(x,z,N)\right].
\label{pf1d}
\end{equation}

The restricted partition functions satisfy a pair of simple recursion relations
\begin{eqnarray}
Z_1(x,z,N) &=& z \int_x^1 Z_0(y,z,N-1)\,dy + \int_x^1 Z_1(y,z,N-1)\, dy \label{recur1} \\
Z_0(x,z,N) &=&  \int_0^x \left[Z_0(y,z,N-1)+Z_1(y,z,N-1)\right]\, dy.
\label{recur2}
\end{eqnarray}
These recursion relations can be easily understood by considering all possibilities
when one adds a new site to the chain. If the new site is a minimum, we need to attach
a factor $z$. On the other hand, if the $(N-1)$-th site was a minimum and it ceases to
be a minimum after the addition of the $N$-th site, we have to detach a factor $z$.

It follows from Eqs. (\ref{recur1}) and (\ref{recur2}) that the restricted partition
functions satisfy the boundary conditions: $Z_1(x=1, z,N)=0$ and $Z_0(x=0,z,N)=0$.
For large $N$, one expects a separation of variables between $x$ and $N$ of the form,
\begin{equation}
Z_{1,0}(x,z,N) \sim \lambda^{-N}\, f_{1,0}(x)
\label{sep1}
\end{equation}
where $\lambda(z)$ is a function of $z$ only (but independent of $x$) and the functions
$f_1(x)$ and $f_0(x)$ satisfy the boundary conditions: $f_1(1)=0$ and $f_0(0)=0$. 
Substituting this ansatz in Eqs. (\ref{recur1}) and (\ref{recur2}) and subsequently
differentiating with respect to $x$, we get a pair of differential equations
\begin{eqnarray}
\frac{df_1}{dx} &=& -\lambda f_1(x) - z\lambda f_0(x) \label{recur3}\\
\frac{df_0}{dx} &=& \lambda f_1(x) + \lambda f_0(x) \label{recur4} 
\end{eqnarray}  
Diagonalizing the $[2\times 2]$ matrix, one obtains the solutions
\begin{eqnarray}
f_1(x) &= & a\, e^{\lambda\, \sqrt{1-z}\, x} + b\, e^{-\lambda\, \sqrt{1-z}\, x} \label{sol1}\\
f_0(x) &=& -\frac{a}{\left(1-\sqrt{1-z}\right)}\, e^{\lambda\, \sqrt{1-z}\, x} - 
\frac{b}{\left(1+\sqrt{1-z}\right)}\, e^{-\lambda\, \sqrt{1-z}\, x} \label{sol2}
\end{eqnarray}
where $a$ and $b$ are arbitrary constants. The two boundary conditions $f_1(1)=0$ 
and $f_0(0)=0$ yield two relations between $a$ and $b$
\begin{eqnarray}
a\, e^{\lambda\, \sqrt{1-z}} + b\, e^{-\lambda\, \sqrt{1-z}}&=& 0 \label{rel1}\\
\frac{a}{\left(1-\sqrt{1-z}\right)}\, e^{\lambda\, \sqrt{1-z}}+\frac{b}{\left(1+\sqrt{1-z}\right)}\, 
e^{-\lambda\, \sqrt{1-z}} &=& 0. \label{rel2}
\end{eqnarray}
Eliminating $a$ and $b$ between Eqs. (\ref{rel1}) and (\ref{rel2}) determines the eigenvalue 
$\lambda(z)$ exactly
\begin{eqnarray}
\lambda(z) & =& \frac{1}{2\sqrt{1-z}}\, \ln \left[\frac{1+\sqrt{1-z}}{1-\sqrt{1-z}}\right],\quad\quad {\rm 
for}\quad\quad 0\le z \le 1 \label{eigen1} \\
&=& \frac{1}{\sqrt{z-1}} \tan^{-1} \left[\sqrt{z-1}\right] \quad\quad {\rm
for}\quad\quad z \ge 1 .\label{eigen2}
\end{eqnarray}
Note that the function $\lambda(z)$ is analytic at $z=1$. The form in Eq. (\ref{eigen2})
is just an analytical continuation of the form in Eq. (\ref{eigen1}) for $z\ge 1$.

Substituting the large $N$ form of $Z_1$ and $Z_0$ in Eq. (\ref{pf1d}), one obtains the large $N$
behavior of the partition function,
\begin{equation}
Z(z,N)= \sum_M P(M,N)\, z^M  \sim [\lambda(z)]^{-N}
\label{pfn1}
\end{equation}
where $\lambda(z)$ is given in Eqs. (\ref{eigen1}) and (\ref{eigen2}). Substituting the anticipated
form of $P(M,N)\sim \exp[-N \Phi(M,N)]$ in Eq. (\ref{pfn1}) one gets
\begin{equation}
Z(z,N) = \sum_M P(M,N)\, z^M \sim \int dy \exp\left[-N\left(\Phi(y)-y \ln z\right)\right]\sim [\lambda(z)]^{-N}.
\label{pfn2}
\end{equation}
Taking the $N\to \infty$ limit in Eq. (\ref{pfn2}) gives
\begin{equation}
{\rm min}_y\left[\Phi(y) - y\, \ln z\right] = \ln (\lambda(z)).
\label{pfn3}
\end{equation}
Inverting via the Legendre transform finally gives the large deviation function
\begin{equation}
\Phi(y) = {\rm max}_z\left[ \ln(\lambda (z)) + y\, \ln z\right],
\label{ld3}
\end{equation}
where $\lambda(z)$ is given in Eqs. (\ref{eigen1}) and (\ref{eigen2}).
Note that determining $\Phi(y)$ from Eq. (\ref{ld3})
requires a knowledge of $\lambda(z)$ for all $z\ge 0$, thus we need both 
formulae of $\lambda(z)$ in Eqs. (\ref{eigen1}) and (\ref{eigen2}). 
\begin{figure}[htbp]
\includegraphics[width=10cm, height=7cm]{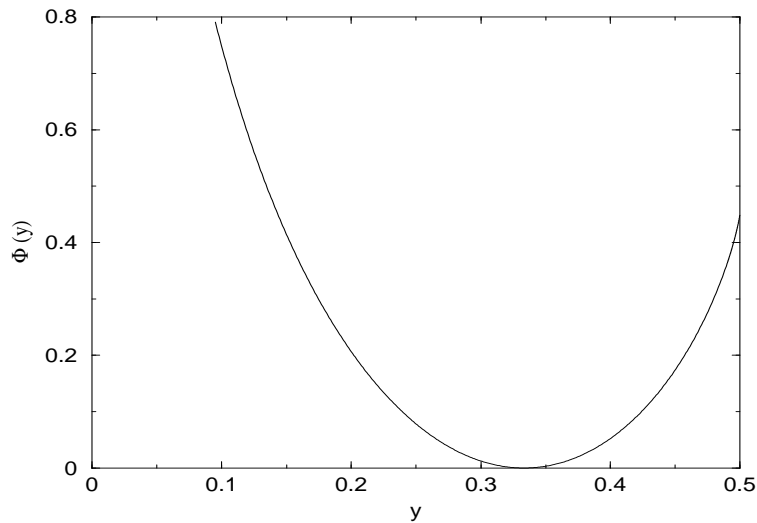}
\caption{The large deviation function $\Phi(y)$ for $0\le y\le 1/2$ in $1$-d
obtained from Eq. (\ref{ld3}) using Mathematica.}
\label{fig:ldf1d}
\end{figure}

We have obtained $\Phi(y)$ from Eq. (\ref{ld3}) using Mathematica and 
it is displayed in Fig.~\ref{fig:ldf1d}.
Since the maximal value of $M$ in $1$-d is $N/2$, the allowed range of $y$ is $0\le y\le 1/2$.
One can analytically obtain the form of $\Phi(y)$ in the three limiting cases $y\to 0$, $y\to 1/3$
and $y\to 1/2$. First let us consider the limit $y\to 0$. To find $\Phi(y)$ in this limit
we need to use the $z\to 0$ form of $\lambda(z)$ in Eq. (\ref{ld3}). As $z\to 0$, it is easy to 
see from Eq. (\ref{eigen1}) that $\lambda(z) \to \ln(4/z)/2$. Substituting this behavior in Eq. 
(\ref{ld3}) and subsequently maximizing the r.h.s of Eq. (\ref{ld3}) we find that $\Phi(y)$
diverges logarithmically,
$\Phi(y) \approx \ln\left[\ln 2/(ey)\right]$ as $y\to 0$. Next we consider the $y\to 1/3$ limit.
Note that the mean number of minima $\langle M \rangle =N/3$ in $d=1$ which follows from
the general result in Eq. (\ref{mean2}). Thus $y\to 1/3$ limit corresponds to
behavior of $M$ near its mean and one expects a quadratic form for $\Phi(y)$ near
$y=1/3$. Indeed this also follows from Eq. (\ref{ld3}). 
The limit $y\to 1/3$ corresponds to using the $z\to 1$ behavior of $\lambda(z)$ in Eq. (\ref{ld3}). 
Substituting $z=1+\epsilon$ in 
Eq.~(\ref{eigen2}) and expanding in powers of $\epsilon$, one gets
$\ln (\lambda(z))= -\epsilon/3 + 13\epsilon^2/90 + O(\epsilon^3)$. 
Substituting this result on the r.h.s of Eq.~(\ref{ld3}) and 
maximizing one gets the expected quadratic
behavior, $\Phi(y)\approx 45(y-1/3)^2/4$ near $y\to 1/3$. This is thus a special
case in $1$-d of the general behavior in Eq.~(\ref{qua}) 
with $a=1/3$ and $b=2/45$.
Finally, to derive the maximally packed limit $y\to 1/2$, we need to use
the $z\to \infty$ behavior of $\lambda(z)$ in Eq.~(\ref{ld3}). In 
this limit, it follows from Eq.~(\ref{eigen2}) 
that $\ln (\lambda(z))\to \ln(\pi/2)-\ln(z)/2-2z^{-1/2}/{\pi}$.
Using this form and maximizing the r.h.s. of Eq. (\ref{ld3}) we get
$\Phi(y) \approx \ln (\pi/2) + 2(1/2-y)\, \ln(\pi(1/2-y)/e)$ 
as $y\to 1/2$. Thus, summarizing the three limiting behaviors
\begin{eqnarray}
\Phi(y) &\approx &  \ln\left[\ln 2/(ey)\right] \quad\quad {\rm as}\quad\quad y\to 0 \nonumber \\
&\approx &  \frac{45}{4}\, (y-1/3)^2 \quad\quad {\rm as}\quad\quad y\to 1/3 \nonumber \\    
&\approx &  \ln (\pi/2) + 2(1/2-y)\, \ln(\pi(1/2-y)/e) \quad\quad {\rm as}\quad\quad y\to 1/2.
\label{limiting}
\end{eqnarray}
Note that as one approaches the maximally packed limit $y\to 1/2$, $\Phi(y_{\rm max}=1/2)=\ln (\pi/2)$. 
Thus, it follows from Eq. (\ref{ld2}) that in $1$-d, for large $N$
\begin{equation}
P(M_{\rm max}=N/2, N) \sim \gamma^{-N}, \quad\quad {\rm with}\quad\, \gamma=\pi/2,
\label{gamma1d}
\end{equation}
the result declared in Eq. (\ref{g1d}).

\section{Probability of the Maximally Packed Configuration: Solvable Cases}
\label{sec:MaxPack}

In this section, we focus only on the maximally packed configuration 
(where the number of minima on the lattice is maximal). From the general 
large deviation theory,
we have already argued that the probability of such a 
configuration $P(M_{\rm max}, N)$
is expected to decay exponentially with the system 
size $N$ as in Eq. (\ref{ld2}).
The goal is to compute the nontrivial constant $\gamma$. In the previous section,
we have shown that in a $1$-d chain, $\gamma=\pi/2$. In this section, we 
compute $\gamma$ exactly in a few other solvable cases, notably for a
Cayley tree with $\mu$ branches and also for a two-leg ladder in $2$-d.

\subsection{Exact Calculation of $\gamma$ on a Cayley Tree}
\label{subsect:Tree}

\begin{figure}[htbp]
\includegraphics[width=12cm]{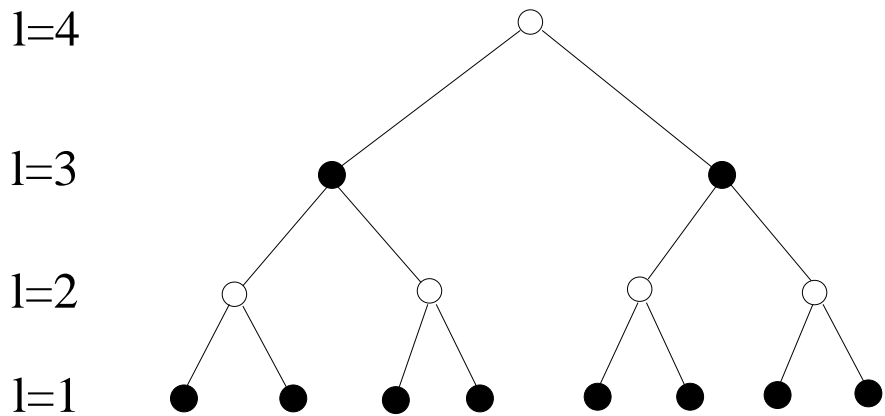}
\caption{The maximally packed configuration of local minima (denoted 
by black dots)
on a Cayley tree with $\mu=2$ branches and $n=4$ generations. The 
layers are labelled
$l=1,2\dots$ starting with the bottom layer.}
\label{fig:tree}
\end{figure}

We consider a Cayley tree with $\mu$ branches and $n$ generations. We label the
generations by $l=1,2,\ldots, n$ starting from the leaf sites at the bottom
(see Fig.~\ref{fig:tree}). The total number of sites on the tree is 
\begin{equation}
N= 1+\mu+\mu^2+\ldots + \mu^{n-1}= \frac{\mu^n-1}{\mu-1}.
\label{nsites}
\end{equation}
Note that in the limit $\mu\to 1^+$, the tree reduces 
to a $1$-d chain with $N= n$ sites.

At each site of this tree resides a random number $x_i$ drawn independently from the
uniform distribution over $[0,1]$. We want to calculate the probability of
the configuration with the maximum possible number of local minima. Note
that if a site is a local minimum, its neighbours cannot be local minima.
To find the
maximally packed configuration, we note that
the number of sites at the bottom layer ($l=1$) increases 
exponentially with $n$ as
$\mu^{n-1}$. Thus, to achieve maximal packing,  it is necessary
to fill up the bottom layer (the leaf sites) with local minima. Then
the layer just above the bottom layer ($l=2$) is devoid of local minima.
The next layer $(l=3$) can again be packed with local minima. Thus, the
maximally packed configuration is the one where alternate layers are fully packed
with minima, starting with a fully packed layer at the bottom, as shown in
Fig.~\ref{fig:tree}. The total number of minima $M_{\rm max}$ in this
maximally packed configuration depends on whether the number of generations
$n$ is even or odd  
\begin{eqnarray}
M_{\rm max} &=& \frac{\mu^{n+1}-1}{\mu^2-1}\quad\quad {\rm even}\quad n  \label{even1} \\
&=& \frac{\mu^{n+1}-\mu}{\mu^2-1}\quad\quad\ {\rm odd}\quad n \label{odd1}
\end{eqnarray} 
For $n$ even, the root is not a minimum whereas for $n$ odd, 
the root is a minimum.
In either case, for $n$ large, the total number of minima in the maximally packed
configuration is proportional to the total number of sites $N$,
\begin{equation}
M_{\rm max} \approx \frac{\mu}{\mu+1} N.
\label{mmax}
\end{equation}
Having identified the maximally packed configuration, we will next compute the
probability of its occurrence and show that for large $N$ 
\begin{equation}
P(M_{\rm max}, N) \sim \gamma^{-N}, \quad\quad {\rm where}\quad\quad \gamma=\frac{1}{\mu+1}\, 
B\left(\frac{1}{\mu+1},\frac{1}{\mu+1}\right).
\label{gct2}
\end{equation}

For simplicity, let us assume that $n$ is even. One can perform 
an identical calculation
when $n$ is odd. For $n$ even, all the even 
layers (labelled by $l=2m$ with $m=1,2,\ldots, n/2$)
are devoid of local minima whereas all the odd layers (labelled by
$l=2m-1$ with $m=1,2,\ldots,n/2$) are fully packed with local minima. Our aim is to 
write a recursion relation. For this, it is convenient to define
two probabilities $P_{2m}(x)$ and $Q_{2m-1}(x)$ defined respectively
for even and odd layers. We define $P_{2m}(x)$ as the probability
that a subtree of $2m$ generations (counted from the bottom of the tree)
has a random variable $x$ at its root and that $x$ is {\em not} a local minimum.
On the other hand, $Q_{2m-1}(x)$ is the probability that a subtree of
$(2m-1)$ generations (again counted from the bottom of the tree) has 
a random variable $x$ at its root and that $x$ is a local minima.
It is then easy to see that they satisfy the recursion relations
\begin{eqnarray}
P_{2m}(x) & =& \left[ \int_0^x Q_{2m-1}(y)\, dy\right]^{\mu} \label{P1} \\
Q_{2m+1}(x) & =& \left[ \int_x^1 P_{2m}(y) \, dy\right]^{\mu} \label{Q1}
\end{eqnarray}
where $m=1,2,\ldots, n/2$. The recursions start with the initial condition
$Q_1(x)=1$. Since $n$ is even, the root of the full tree with $n$ generations is
not a local minima and hence the probability of the full tree
is just $P_n(x)$ given that the value at the root 
is $x$.  The probability of the maximally packed 
configuration is then obtained by integrating over $x$ at the root
\begin{equation}
P(M_{\rm max},N)= \int_0^1 P_n(x) dx.
\label{pmp1}
\end{equation}
Thus, we need to solve Eqs. (\ref{P1}) and (\ref{Q1}) and then substitute the solution
for $P_n(x)$ in Eq. (\ref{pmp1}) to calculate $P(M_{\rm max},N)$.

To solve the nonlinear recursion relations, it is convenient to define
$p_{2m}(x)= [P_{2m}(x)]^{1/\mu}$ and $q_{2m+1}(x)=[Q_{2m+1}(x)]^{1/\mu}$.
Then the recursions in Eqs. (\ref{P1}) and (\ref{Q1}) become
\begin{eqnarray}
p_{2m}(x)&=& \int_0^x q_{2m-1}^\mu (y)\, dy  \label{P2} \\   
q_{2m+1}(x)&=& \int_x^1 p_{2m}^{\mu}(y)\, dy \label{Q2}
\end{eqnarray}
starting with $q_1(x)=1$ for $0\le x\le 1$.
It follows from Eqs. (\ref{P2}) and (\ref{Q2}) that they satisfy the boundary conditions,
$p_{2m}(0)=0$ and $q_{2m+1}(1)=0$ for all $m\ge 1$. However, $p_{2m}(1)$ and $q_{2m+1}(0)$
are nonzero. It is then useful to define the ratios, $f_{2m}(x)= p_{2m}(x)/p_{2m}(1)$
and $g_{2m+1}(x)= q_{2m+1}(x)/q_{2m+1}(0)$ so that $f_{2m}(1)=1$, $f_{2m}(0)=0$
and $g_{2m+1}(0)=1$, $g_{2m+1}(1)=0$. In terms of the new functions, 
the recursions become
\begin{eqnarray}
f_{2m}(x)&=& \frac{q_{2m-1}^\mu(0)}{p_{2m}(1)}\, \int_0^x g_{2m-1}^\mu (y)\, dy  \label{f1} \\
g_{2m+1}(x)&=& \frac{p_{2m}^\mu(1)}{q_{2m+1}(0)}\, \int_x^1 f_{2m}^{\mu}(y)\, dy \label{g1}
\end{eqnarray}
starting with $g_{1}(x)=1$ for $0\le x\le 1$.
As $m$ increases, we expect that the ratio functions $f_{2m}(x)$ and $g_{2m+1}(x)$ 
will approach their respective $m$-independent fixed point forms $f(x)$ and $g(x)$.
This means that as $m\to \infty$, $\frac{q_{2m-1}^\mu(0)}{p_{2m}(1)}\to \lambda_1$
and $\frac{p_{2m}^\mu(1)}{q_{2m+1}(0)}\to \lambda_2$ and
\begin{eqnarray}
f(x)& =& \lambda_1 \int_0^x g^\mu(y)\, dy \label{f2} \\
g(x)& =& \lambda_2 \int_x^1 f^{\mu}(y)\, dy \label{g2}
\end{eqnarray}
with the boundary conditions $f(0)=0$, $f(1)=1$ and $g(0)=1$, $g(1)=0$. These boundary conditions will
determine the eigenvalues $\lambda_1$ and $\lambda_2$.

Differentiating Eqs. (\ref{f2}) and (\ref{g2}) with respect to $x$ gives
\begin{eqnarray}
\frac{df}{dx} = \lambda_1 g^\mu(x) \label{f3} \\
\frac{dg}{dx}= -\lambda_2 f^\mu(x) \label{g3}.
\end{eqnarray}
Multiplying (\ref{f3}) by $\lambda_2 f^{\mu}(x)$ and (\ref{g3}) by $\lambda_1 g^{\mu}(x)$,
adding and then integrating, we find a conserved 
quantity
\begin{equation}
\lambda_2 f^{\mu+1}(x) + \lambda_1 g^{\mu+1}(x)= C
\label{com1}
\end{equation}
where $C$ is a constant independent of $x$. Putting $x=0, 1$ and using the respective boundary conditions
gives $C=\lambda_1=\lambda_2$. Thus, $\lambda_1=\lambda_2=\lambda$ and
\begin{equation}
f^{\mu+1}(x) + g^{\mu+1}(x) =1.
\label{com2}
\end{equation}
The common eigenvalue $\lambda$ is yet to be determined.
Eliminating $g(x)$ between Eqs. (\ref{f3}) and (\ref{com2}) gives
\begin{equation}
\frac{df}{dx} = \lambda \left[1-f^{\mu+1}(x)\right]^{\mu/(\mu+1)}
\label{f4}
\end{equation}
subject to the boundary conditions $f(0)=0$ and $f(1)=1$. 
Integrating and using $f(0)=0$
we get
\begin{equation}
\int_0^{f(x)} \frac{dz}{(1-z^{\mu+1})^{\mu/(\mu+1)}} = \lambda x.
\label{f5}
\end{equation}
Using the other boundary condition $f(1)=1$ determines $\lambda$ explicitly
\begin{equation}
\lambda= \frac{1}{\mu+1}\, B\left(\frac{1}{\mu+1},\frac{1}{\mu+1}\right)
\label{lambda1}
\end{equation}
where $B(m,n)=\int_0^1 x^{m-1}(1-x)^{n-1}dx$ is the standard Beta function.
The eigenfunction $f(x)$ in Eq. (\ref{f5}) can be expressed as the solution
of the equation
\begin{equation}
f(x)\, F\left[\frac{1}{\mu+1}, \frac{\mu}{\mu+1}, \frac{\mu+2}{\mu+1}, f^{\mu+1}(x)\right]
= \frac{x}{\mu+1}\, B\left(\frac{1}{\mu+1},\frac{1}{\mu+1}\right)
\label{f6}
\end{equation}
where $F[a,b,c,z]$ is the hypergeometric function. 
The other eigenfunction $g(x)$ follows from Eq. (\ref{com2}), or simply
from the symmetry $g(x)=f(1-x)$.
To check that the solutions
to the original recursion relations (\ref{P2}) and (\ref{Q2}) indeed converge
to these fixed solutions, we have numerically 
solved Eqs. (\ref{P2}) and (\ref{Q2})
for $\mu=2$. We find that the numerical solutions (the ratio functions
$f_{2m}(x)$ and $g_{2m+1}(x)$) converge to
fixed point functions rather quickly after about $3$ or $4$ iterations.
In Fig.~\ref{f3}, we compare the numerical 
fixed point solution (the solution after $10$ iterations) 
$f(x)$ with the analytical solution in Eq. (\ref{f6}) with $\mu=2$. The agreement is perfect.
\begin{figure}[htbp]
\includegraphics[width=10cm, height=7cm]{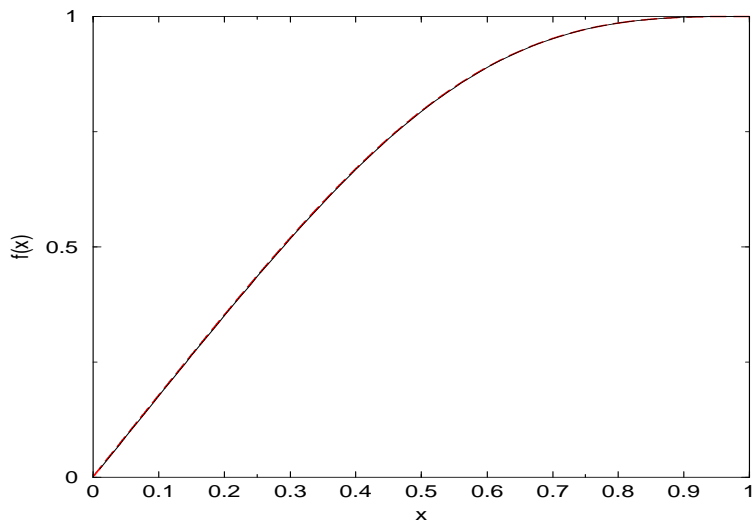}
\caption{The numerically obtained fixed point 
function $f(x)$ after $10$ iterations,
for $\mu=2$,
is compared to the analytical scaling function in Eq. (\ref{f6}) obtained
using Mathematica. The two curves are difficult to distinguish indicating perfect agreement.}
\label{fig:f2}
\end{figure}

To determine $\gamma$, we use Eq.~(\ref{pmp1}) and the 
recursion relation (\ref{Q2}) which show that
\begin{equation}
P(M_{\rm max},N)= \int_0^1 P_n(x) dx= q_{n+1}(0).
\label{pmp2}
\end{equation}
Then we take ratios to extract the bulk contribution
of the Cayley tree, thereby removing artifacts coming from
its surface~\cite{Baxter89}: as $m \to \infty$, we have
$\frac{q_{2m-1}^\mu(0)}{p_{2m}(1)}\to \lambda$
and $\frac{p_{2m}^\mu(1)}{q_{2m+1}(0)}\to \lambda$. Eliminating
$p_{2m}(1)$ gives a recursion for large $m$
\begin{equation}
q_{2m+1}(0)= \frac{1}{\lambda^{\mu+1}}\, q_{2m-1}^{\mu^2} (0).
\label{q1}
\end{equation}
Iterating it, we obtain, for large $m$, $q_{2m+1}(0)\sim \lambda^{-(\mu^{2m}-1)/(\mu-1)}$.
Substituting this result in Eq. (\ref{pmp2}) and 
using $N=(\mu^n-1)/(\mu-1)$ gives our final result for large $N$
\begin{equation}
P(M_{\rm max},N)\sim \gamma^{-N} \quad\quad 
{\rm with}\quad \gamma=\lambda=\frac{1}{\mu+1}\,
B\left(\frac{1}{\mu+1},\frac{1}{\mu+1}\right).
\label{pmp3}
\end{equation}

Note that the recursion relation (\ref{q1}) is valid only for large $m$. But in deriving the
results above we have assumed that it holds even for small $m$. Even though the
asymptotic result is not expected to change due to this `initial condition' effect, one
can avoid this ``surface effect'' by 
appropriately defining $\gamma$ as the ratio
\begin{equation}
\frac{1}{\gamma}= \lim_{n\to \infty} 
\frac{\int_0^1 \left[\int_x^{1} P_n(y)dy\right]^{\mu} dx}
{\left[\int_0^1 P_n(x)dx\right]^{\mu}}.
\label{gnf1}
\end{equation}
This definition follows from the following observation. 
The numerator on the r.h.s of (\ref{gnf1}) denotes the probability
of a maximally packed configuration on a tree with $(n+1)$ generations.
The denominator is the joint probability that the disconnected $\mu$ subtrees
of $n$ generations are all maximally packed. Thus, the ratio on the r.h.s.
is just the factor by which the probability of a 
maximally packed configuration per site
decreases when one fuses the $\mu$ number of $n$-generation trees 
and one additional
root to construct a newly maximally packed configuration on 
a $(n+1)$-generation tree.
But, asymptotically for large $n$, this is precisely $1/\gamma$ by the 
orginal definition in Eq. (\ref{ld2}). Physically,
$\ln \gamma$ is the additional entropy change when packing
an extra minimum in the tree. 
Using the recursion relations (\ref{P1}), (\ref{Q1}) in Eq. (\ref{gnf1}) one gets
\begin{equation}
\frac{1}{\gamma}=\lim_{n\to \infty} \int_0^1 \left[\frac{q_{n+1}(x)}{q_{n+1}(0)}\right]^{\mu} dx=\int_0^1  
g_{n+1}^\mu(x) dx.
\label{gnf2}
\end{equation}
Substituting $g^{\mu}(x)= \frac{1}{\lambda}\,df/dx$, integrating and using $f(1)=1$, we get
\begin{equation}
\gamma= \lambda=\frac{1}{\mu+1}\,
B\left(\frac{1}{\mu+1},\frac{1}{\mu+1}\right).
\label{gnf4}
\end{equation}

Note that for $\mu=1$, we recover the $1$-d result, $\gamma=\pi/2$. For $\mu=2$, we get
$\gamma= B(1/3,1/3)/3=1.76664\ldots$. As $\mu\to \infty$, $\gamma$ converges slowly
to $\gamma=2$. Since the $\mu\to \infty$ result should coincide with that
on a hypercubic lattice in the infinite dimension limit, we expect that for
hypercubic lattices in $d$ dimensions, $\gamma$ is bounded, 
$\pi/2\le \gamma \le 2$.
As $d$ increases from $1$ to $\infty$, $\gamma$ should increase monotonically
from its $d=1$ value $\pi/2=1.56079\ldots$ to $2$. Our numerical solutions
on $d$-dimensional hypercubic lattices with $d=2,3,4,5$ are consistent
with these bounds.

\subsection{Maximally Packed Configuration on a Bipartite 
lattice: An Equivalent Plaquette Model}
\label{subsect:Plaquette}

\begin{figure}[htbp]
\includegraphics[width=10cm]{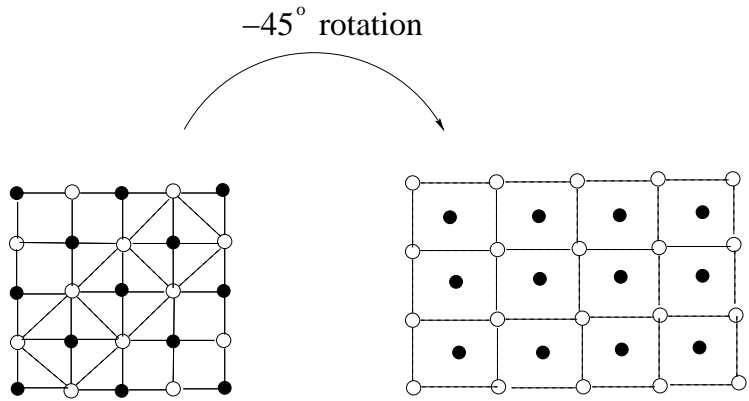}
\caption{A maximally packed configuration on a square lattice. On this checkerboard pattern, the black circles
denote the location
of the local minima. The corners of every plaquette around a minima 
cannot contain a
minimum and are shown by empty circles. On the right, the 
same pattern, rotated by an angle $-45^o$.}
\label{fig:plaque}
\end{figure}

On a bipartite lattice, the maximally packed configuration is the one where
one places a local minimum at every alternate site. For example, on a square lattice,
a maximally packed configuration has a checkerboard pattern as shown 
in Fig.~\ref{fig:plaque}.
Let us first focus on the neighbourhood of a single local minimum. The sites at the
corners of the plaquette containing this minimum are clearly not local minima. Let
$x_1$, $x_2$, $x_3$ and $x_4$ denote the values of the random variables at the four
corners of the plaquette. Then, given these four values, the probability that
the site at the center of the plaquette is a local minimum is clearly
\begin{equation}
p_i(x_1,x_2,x_3,x_4)= {\rm min}\,(x_1,x_2,x_3,x_4)
\label{plaq1}
\end{equation}
where we have used the fact that the random variable 
at the center of the plaquette
is drawn from a uniform distribution. So, the probability of 
the full checkerboard
configuration where every alternate site is a local minimum, given the 
values of the random variables $\{x_i\}$'s at 
the sites of the other sublattice, is obtained 
by multiplying all the plaquettes
\begin{equation}
{\rm Prob}\left(M_{\rm max}, N\,|\,\{x_i\}\right)= \prod_{{\rm plaquette}\, j} {\rm min}\left(x_1(j),x_2(j), 
x_3(j), 
x_4(j)\right)
\label{plaq2}
\end{equation}
where the plaquettes are labelled by $j$ and $x_1(j)$, $x_2(j)$, $x_3(j)$ and $x_4(j)$ are
the four random variables at the corners of the $j$-th plaquette. 
The total number of plaquettes will also be denoted by $M$ since the number of plaquettes
is the same as the number of local minima.
To simplify, it is
convenient to rotate the lattice by $-45^o$ (as shown in Fig.~\ref{fig:plaque}).

We then have a plaquette model where at each corner of a plaquette lives a random variable $x_i$
drawn from a uniform distribution over $[0,1]$ and we are interested in calculating
the product in Eq. (\ref{plaq2}). Finally, the probability of the 
maximally packed configuration
on the original lattice is obtained by averaging 
over the $x_i$'s with uniform distribution
in $[0,1]$
\begin{equation}
P(M_{\rm max}, N) = \big \langle \prod_{{\rm plaquette}\, j} {\rm min}\left(x_1(j),x_2(j), x_3(j),
x_4(j) \right) \big \rangle 
\label{plaq3}
\end{equation}
where the angled brackets simply indicates integration over all the $x$ variables from $0$ to $1$.
Note that this plaquette model is very general 
and can be extended to any bipartite
lattice. Also, the right hand
side of Eq.~(\ref{plaq3}) is actually the probability of just one of the two
checkerboard configurations, so the true $P(M_{\rm max}, N)$
is actually twice this amount, but to avoid unnecessary complications
we shall keep to this notation.

\subsubsection{Plaquette Model in One Dimension}

As a simple example, for a $1$-d chain, one can reproduce the result $\gamma=\pi/2$ quite easily
using the plaquette representation. The number of plaquettes is clearly $M=N/2$.
In this case, Eq.~(\ref{plaq3}) 
gives
\begin{equation}
P(M_{\rm max}, N) =\big\langle \prod_{i=1}^{M=N/2} {\rm min} (x_i, x_{i+1}) \big\rangle
\label{cd1}
\end{equation}
where $i$ runs over every alternate sites of the original $1$-d chain. The quantity 
on the r.h.s. of (\ref{cd1}) can be evaluated using a simple transfer matrix approach. 
Defining a transfer matrix via, 
$\langle x_i|{\hat T}| x_{i+1}\rangle = {\rm min}(x_i, x_{i+1})$, 
we have from Eq. (\ref{cd1}), assuming a closed periodic chain
\begin{equation}
P(M_{\rm max}, N)= {\rm Tr}\left[ {\hat T}^M\right] 
\label{cd2}
\end{equation}
where ${\rm Tr}$ is the trace. The eigenvalue equation of the transfer matrix is 
\begin{equation}
\int_0^{1} {\rm min}(x,y)\, \psi(y)\, dy = \lambda\, \psi(x)
\label{cd3}
\end{equation}
where $\psi(x)$ is an eigenfunction with eigenvalue $\lambda$. Dividing the range of integration
into $[0,x]$ and $[x,1]$ and differentiating twice, one gets 
\begin{equation}
\lambda \frac{d^2 \psi}{dx^2}+ \psi(x)=0
\label{cd4}
\end{equation}
with the boundary conditions $\psi(x=0)=0$ and $\psi'(x=1)=0$ where $\psi'(x)=d\psi/dx$.
The solution
is simply, $\psi(x)= A \sin(x/\sqrt{\lambda})$ where 
\begin{equation}
\lambda= \frac{4}{\pi^2 (2m+1)^2}; \quad\quad m=0,1,2\ldots 
\label{cd5}
\end{equation}
The largest eigenvalue is $\lambda=4/\pi^2$ corresponding to $m=0$. Thus for large
$M=N/2$, one gets from Eq. (\ref{cd2})
\begin{equation}
P(M_{\rm max}, N)\sim \left[\frac{4}{\pi^2}\right]^{N/2}\sim [\pi/2]^{-N}
\label{cd6}
\end{equation}
reproducing the results obtained in the previous section.

\subsubsection{Plaquette Model on a Two-leg Ladder}

\begin{figure}[htbp]
\includegraphics[width=12cm]{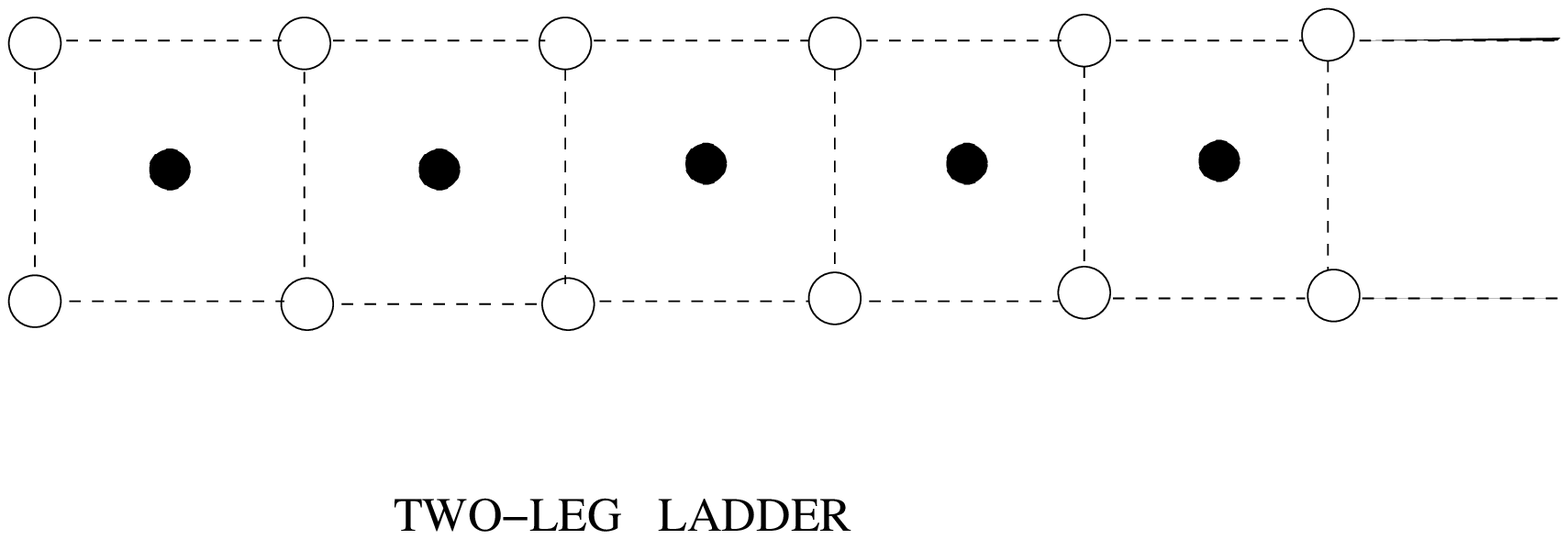}
\caption{The plaquette model on a two-leg ladder. At the center of each plaquette
lives a local minimum shown by the black circles.}
\label{fig:ladder}
\end{figure}

Another nontrivial solvable case of the plaquette model is on 
a two-leg ladder shown in Fig.~\ref{fig:ladder}.
It can be thought of as the first layer of the full $2$-d 
model shown on the right
in Fig.~\ref{fig:plaque}. At the center of each plaquette lives a local minimum.
Let $M$ be the number of plaquettes. Then the total number of sites in the lattice
(counting the minima at the centers) is $N=3M$. 

The probability of the maximally
packed configuration in Eq. (\ref{plaq2}) can again be computed by defining
the transfer matrix, $\langle x_1, x_2|{\hat T}| x_3, x_4\rangle= {\rm min}(x_1,x_2,x_3,x_4)$
where $(x_1,x_2)$ refers to the two corners on the left of a plaquette and
$(x_3,x_4)$ refers to the two corners on the right of a plaquette. 
Then, 
\begin{equation}
P(M_{\rm max}, N)= {\rm Tr}\left[ {\hat T}^M\right]
\label{l1}
\end{equation}
The corresponding
eigenvalue equation is now two dimensional
\begin{equation}
\int_0^1 dx_3\, \int_0^1\, dx_4\, {\rm min}(x_1,x_2,x_3,x_4)\, \psi(x_3,x_4) = \lambda\, \psi(x_1,x_2)
\label{l2}
\end{equation}
which is considerably harder to solve compared to the $1$-d case. While
one can solve the $2$-d eigenvalue equation directly, it is
somewhat easier to first reduce it to an equivalent $1$-d eigenvalue problem 
by using the following trick.

\begin{figure}[htbp]
\includegraphics[width=14cm]{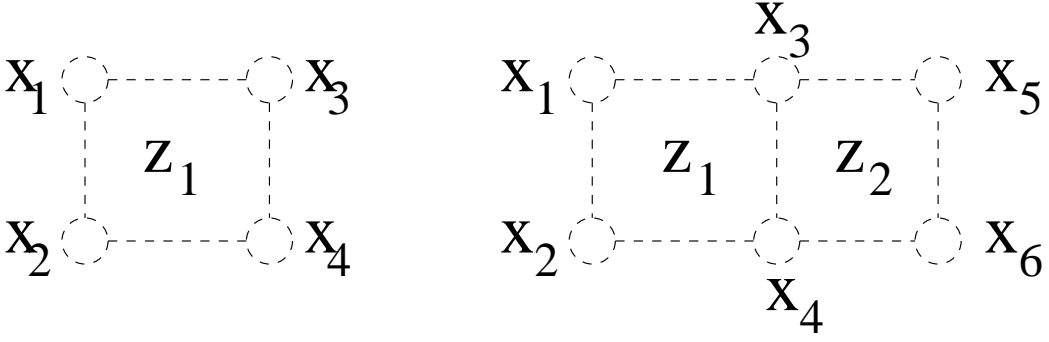}
\caption{On the left we have a single plaquette with the minimum $z_1={\rm min}(x_1,x_2,x_3,x_4)$
at the center. On the right we have two adjacent plaquettes with minima
$z_1={\rm min}(x_1,x_2,x_3,x_4)$ and $z_2={\rm min}(x_3,x_4,x_5,x_6)$ at their respective centers. The variables
$z_1$ and $z_2$
are corrrelated as they share the common bond with the two elements
$x_3$ and $x_4$.}
\label{fig:ladder1}
\end{figure}

Let us first define the random variable $z_i={\rm min}(x_1(i),x_2(i), x_3(i), x_4(i))$ at the
center of each plaquette $i$. Clearly
\begin{equation}
P(M_{\rm max}, N)= \langle z_1 z_2\ldots z_M\rangle
\label{l3}
\end{equation}
where $M=N/3$ is the number of plaquettes. 
The random variables $z_i$'s are obviously correlated as two adjacent plaquettes
will share two common random variables as shown in Fig.~\ref{fig:ladder1}.

To evaluate the average on the r.h.s
of Eq. (\ref{l3}) we need to know the joint distribution of the $z_i$'s
which are correlated (see Fig.~\ref{fig:ladder1}). 
This joint distribution can be explicitly computed.
To see this, let us first compute the cumulative distribution of one single
$z$ variable, say $z_1={\rm min}(x_1(1), x_2(1), x_3(1), x_4(1))$.
For simplicity, we denote $x_1(1)=x_1$, $x_2(1)=x_2$ etc. (see 
Fig.~\ref{fig:ladder1}).
Evidently
\begin{equation}
{\rm Prob}(z_1>y) = {\rm Prob}({\rm min}(x_1,x_2,x_3,x_4)>y)=(1-y)^4  
\label{l4}
\end{equation}
where we have used the fact that $x_1$, $x_2$, $x_3$ and $x_4$ are all independent
random variables drawn from the uniform distribution over $[0,1]$. Next, let us consider
the joint distribution of two consecutive $z$'s, say $z_1$ and $z_2$. Let us
denote the random variables at the corners by $x_1$, $x_2$, $x_3$, $x_4$, $x_5$
and $x_6$ as shown in Fig.~\ref{fig:ladder1}. Then it is easy to see that 
\begin{eqnarray}
{\rm Prob}(z_1>y_1, \,z_2>y_2)&= &{\rm Prob}({\rm min}(x_1,x_2,x_3,x_4)>y_1,\, {\rm min}(x_3,x_4,x_5,x_6)>y_2) 
\nonumber \\
&=& (1-y_1)^2\, \left[1-{\rm max}(y_1,y_2)\right]^2\,(1-y_2)^2.
\label{l5}
\end{eqnarray}
In a similar way, one can construct the joint distribution of three adjacent plaquette minima
\begin{equation}
{\rm Prob}(z_1>y_1,\, z_2>y_2,\, z_3>y_3)= (1-y_1)^2\, \left[1-{\rm max}(y_1,y_2)\right]^2\,
\left[1-{\rm max}(y_2,y_3)\right]^2\, (1-y_3)^2.
\label{l6}
\end{equation}
One can repeat the process above for higher number of adjacent 
plaquettes and one immediately sees the pattern
for the full ladder. Let us denote 
\begin{equation}
F(y_1,y_2,y_3,\ldots, y_M)={\rm Prob}(z_1>y_1,\, z_2>y_2,\, z_3>y_3,\, \ldots, z_M>y_M).
\label{fd}
\end{equation}
Then, for a ladder with open ends, we get
\begin{equation}
F(y_1,y_2,y_3,\ldots, y_M)= (1-y_1)^2\,\left[ \prod_{i=2}^{M}\left[1-{\rm 
max}(y_{i-1},y_i)\right]^2\,\right] (1-y_M)^2.
\label{l7}
\end{equation}
For a ladder with closed (periodic) ends, this is even simpler
\begin{equation}
F(y_1,y_2,y_3,\ldots, y_M)= \prod_{i=1}^{M}\left[1-{\rm
max}(y_{i-1},y_i)\right]^2.
\label{l8}
\end{equation}
where one identifies $y_0=y_M$.

Once we have the joint distribution, it is easy to rewrite the average on the r.h.s. of 
Eq. (\ref{l3}) in terms of the joint distribution
\begin{equation}
P(M_{\rm max}, N)= \big \langle z_1 z_2\ldots z_M\big \rangle= \int_{0}^1\ldots \int_0^1 dy_1 dy_2\ldots dy_M 
\, F(y_1,y_2, y_3,\ldots, y_M)  
\label{l9}
\end{equation}
The latter identity can be easily derived using integration by parts.
Let us, for simplicity, consider a plaquette with closed ends. Then, substituting
the joint distribution from Eq. (\ref{l8}) into Eq. (\ref{l9}), we now have
a one-dimensional multiple integral to peform 
\begin{equation}
P(M_{\rm max}, N)=\int_{0}^1\ldots \int_0^1 dy_1 dy_2\ldots dy_M\,  \prod_{i=1}^{M}\left[1-{\rm
max}(y_{i-1},y_i)\right]^2.
\label{l10}
\end{equation}
Using the identity $1-{\rm max}(y_1,y_2)= {\rm min}(1-y_1,1-y_2)$ and making a change of variable
$x_i=1-y_i$, the integral in Eq. (\ref{l10}) simplifies further 
\begin{equation}
P(M_{\rm max}, N)= \int_{0}^1\ldots \int_0^1 dx_1 dx_2\ldots dx_M \,\prod_{i=1}^{M}\left[{\rm 
min}^2(x_{i-1},x_i)\right].
\label{l11}
\end{equation}
This one dimensional integral can now be performed by a transfer matrix technique which
operates in one dimension. Defining $\langle x_{i-1}|{\hat T}|x_i \rangle = {\rm min}^2(x_{i-1}, x_i)$
we get, $P(M_{\rm max}, N)={\rm Tr}\left[{\hat T}^M\right]$ and the eigenvalue equation is given by
\begin{equation}
\int_0^1 {\rm min}^2(x,y)\, \psi(y)\, dy= \lambda \,\psi(x).
\label{l12}
\end{equation}
Thus we have managed to reduce a two-dimensional eigenvalue problem in Eq. (\ref{l2}) into
an equivalent but much simpler $1$-d eigenvalue problem in Eq. (\ref{l12}) which can
then be solved exactly.

To proceed, we first divide the range of integration in Eq. (\ref{l12}) into $[0,x]$ and $[x,1]$.
Next we differentiate once to get
\begin{equation}
\lambda \frac{d\psi}{dx}= 2x\, \int_x^1 \psi(y) dy.
\label{l13}
\end{equation}
Note the boundary conditions that emerge from Eqs.(\ref{l12}) and (\ref{l13}): $\psi(x=0)=0$
and $\psi'(x=0)=0$. But in addition one also has to satisfy $\psi'(x=1)=0$. 
Dividing Eq. (\ref{l13}) by $x$ and 
differentiating
once more we get an ordinary second order differential equation
\begin{equation}
\psi''(x) -\frac{1}{x} \psi'(x) +\frac{2x}{\lambda}\, \psi(x)=0
\label{l14}
\end{equation}
with the boundary conditions: $\psi(x=0)=0$,
$\psi'(x=0)=0$ and $\psi'(x=1)=0$. The general solution to this equation, after a few changes of variables,
can fortunately be obtained explicitly as a linear combination of two independent Bessel functions. 
The boundary condition
$\psi(0)=0$ rules one of them out. Then, the most general solution satisfying
$\psi(0)=0$ can be written as
\begin{equation} 
\psi(x)= A\, x\, J_{2/3}\left(\sqrt{\frac{8x^3}{9\lambda}}\right)
\label{l15}
\end{equation}
where $J_{\nu}(x)$ is the ordinary Bessel function with 
index $\nu$~\cite{GradshteynRyzhik65}
and $A$ is an arbitrary amplitude. Note that the other boundary condition at $x=0$, namely
$\psi'(0)=0$ is automatically satisfied by the solution in Eq. (\ref{l15}). This can be seen by using
the small $x$ expansion of $J_{\nu}(x) \sim x^{\nu}$ which indicates $\psi(x)\sim x^2$ as $x\to 0$.
Hence $\psi'(0)=0$. 

To determine the eigenvalue $\lambda$, we have to use the other
nontrivial boundary condition at $x=1$, namely $\psi'(x=1)=0$.
This condition, substituted in Eq. (\ref{l15}), gives us an implicit equation 
for $\lambda$ that looks a bit complicated
\begin{equation}
\frac{2}{3} J_{2/3}\left(\sqrt{\frac{8}{9\lambda}}\right)+\sqrt{\frac{8}{9\lambda}}\, 
J'_{2/3}\left(\sqrt{\frac{8}{9\lambda}}\right)=0.
\label{l16}
\end{equation}
However, a nice simplication occurs when one uses the 
identity~\cite{GradshteynRyzhik65}, $x J'_{\nu}(x)+\nu J_{\nu}(x)=x 
J_{\nu-1}(x)$. Then Eq. (\ref{l16}) simply gives
\begin{equation}
J_{-1/3}\left(\sqrt{\frac{8}{9\lambda}}\right)=0.
\label{l17}
\end{equation}
The Bessel function oscillates on the positive axis of its argument, so each of its zeroes
(roots) would give an eigenvalue $\lambda$. However, for large $M$, we are interested only in the
largest eigenvalue which is then given by
\begin{equation}
\lambda= \frac{8}{9\alpha^2}
\label{l18}
\end{equation}
where $\alpha$ is smallest positive root of $J_{-1/3}(x)=0$. The root $\alpha$
is known~\cite{AbramowitzStegun64}, $\alpha=1.86635\ldots$. 

Using $P(M_{\rm max},N)= {\rm Tr}\left[{\hat T}^M\right]\sim \lambda^M$ for large $M$
and using $M=N/3$, we finally get the exact result for the two-leg
ladder
\begin{equation}
P(M_{\rm max},N) \sim \gamma^{-N}; \quad\quad{\rm where}\quad 
\gamma=\left[\frac{9\alpha^2}{8}\right]^{1/3}=1.57657\ldots
\label{l19}
\end{equation}
When compared with the $1$-d result, $\gamma=\pi/2=1.57079\ldots$, we see that $\gamma$
changes by a very small amount as one goes from a chain to a ladder.
It would be interesting to see if one can extend these calculations 
to ladders with
more than two legs~\cite{EvansMajumdar07} and eventually 
to the full $2$-dimensional lattice.

\section{Numerical Results for Maximally Packed $d$-Dimensional Lattices}
\label{sec:Numerics}

Now we consider the dependence of $\gamma$ on dimension. For
simplicity, we have focused on $d$-dimensional hypercubic lattices as 
these lattices are simple to parametrize and are
bipartite, allowing us to use the $d$-dimensional analog of the framework
described in Sect.~\ref{subsect:Plaquette}. 
Our approach will be computational: we numerically estimate
$P(M_{\rm max},N)$ from an integral representation 
for lattices of increasing size 
and then try to extract the limit of large sizes. To 
avoid boundary effects which slow down the convergence to 
this limit, we have used $L^d$ lattices having periodic boundary
conditions in all directions.

The integral in Eq.~(\ref{plaq3}) is based on decomposing the lattice
into even and odd sites and imposing the local minima to be
say on the even sites.
Its generalization to $d$ dimensions involves all the neighbors
of a given even site $j$, the corresponding ``star'' set of
$2d$ odd sites playing the role the plaquette had for the
square lattice:
\begin{equation}
P(M_{\rm max}, N) = \big \langle \prod_{{\rm star}\, j} 
{\rm min}\left(x_1(j),x_2(j), \ldots x_{2d}(j) \right) \big \rangle \, .
\label{eq:star}
\end{equation}
Here the average is over all values of the random variables
belonging to stars; these are uniform i.i.d. in $[0,1]$.
Furthermore, as before, we neglect the factor 2 in 
$P(M_{\rm max}, N)$ coming from the fact that the minima
could have been taken to be on the odd sites.
The difficulty in computing 
these integrals is their high dimensionality. 
In Monte Carlo as
used in most statistical physics applications, it is straightforward
to use importance sampling methods (e.g., the Metropolis algorithm)
to get expectation values of observables; unfortunately here, 
the quantity to compute is the analog of the free energy and
it is not directly accessible via such methods. We have thus used
a different approach that is 
of the ``variance reduction'' type. It can be motivated
as follows. In Eq.~(\ref{eq:star}) we are to get the mean value of the
integrand, sampling the random variables $x_i$ uniformly. It is quite
easy to see that the signal to noise ratio goes to zero 
exponentially with the number of lattice sites; to 
counteract this, we sample the
$x_i$ with a different density and then correct for this biased
sampling. For this to be practical, we keep the $x_i$
as i.i.d. variables, but optimize their individual distribution
so as to maximize the signal to noise ratio.
Let $\rho(x)$ be the probability density used for sampling
the $x_i$. For any such distribution, 
\begin{equation}
P(M_{\rm max}, N) = \big \langle \prod_{{\rm star}\, j} 
\frac{{\rm min}\left(x_1(j),x_2(j), \ldots x_{2d}(j) \right)}
{\left[ \rho(x_1(j)) \rho(x_2(j)) \ldots \rho(x_{2d}(j)) \right]^{1/2d} }
\big \rangle_{\rho} .
\label{eq:starVR}
\end{equation}
The denominator corrects for the modified measure of the random
variables, taking into account the fact that each odd site
appears in $2d$ stars. If $\rho$ is well chosen, the 
numerator and denominator of the integrand will fluctuate
together so that their ratio has a reduced variance. For our purposes,
we parametrized $\rho$ as follows:
\begin{eqnarray}
\rho(x) &=& A x^{\phi} \quad\quad x \le x^* \nonumber \\
\rho(x) &=& B \quad\quad\quad x \ge x^*
\end{eqnarray} 
where $\phi$ and $x^*$ are arbitrary parameters while $A$ and $B$ are 
set so that $\rho$ is continuous and is a normalized probability density.
For each dimension, we adjust $\phi$ and $x^*$ to minimize
the variance the integrand; the signal to noise ratio
still decreases exponentially with the number of lattice
sites, but with a smaller exponent. The motivation for this functional
form is simple: the odd sites, all of which are maxima,
have an a posteriori distribution that is strongly suppressed
at low values of the random variable. 

One last obstacle
comes from the fact that the integrand typically
takes on small values when one takes large lattices; this
is expected of course since the integral itself is 
becoming exponentially small. To keep track of values
that are smaller than what can be represented by the
machine coding, (we used 96 bit representations
of real numbers), we shifted multiplicatively each star term
in the integrand and corrected for this shift when computing
$\ln(P(M_{\rm max},N))$. The set of these procedures then gave
us values of $\ln(P(M_{\rm max},N))/N$ with measurable statistical
errors for a range of $N=L^d$ from which we extrapolated to the
large $N$ limit.

\begin{figure}[htbp]
\includegraphics[width=10cm]{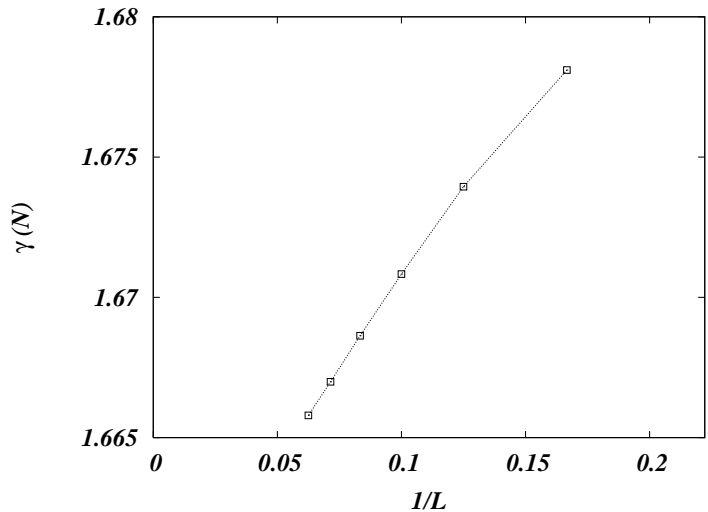}
\caption{The convergence to the large $L$ limit of the 
estimates of $\gamma(N)$ for the square lattice, $N=L^2$.}
\label{fig:gammaConvergence}
\end{figure}

In practice, we find that this strategy works very well in low 
dimensions. For instance in dimension $d=1$, we still have a 
very good precision for $L=30$ sites, and the large $L$ limit can 
be very reliably extracted, giving $\gamma$ to better than 5 
significant figures. (Of course, since the exact value is known at $d=1$, 
this really only serves
as a check of our procedures.)
In dimension $d=2$, the method gives better than 4 significant figures
for $L \approx 16$. Our estimates of $\gamma(N) = P(M_{\rm max},N)^{-1/N}$
as a function of $1/L=1/\sqrt{N}$ are shown in 
Fig.~(\ref{fig:gammaConvergence}).
The convergence to the $L=\infty$ limit
seems to follow a $1/L$ law, a property we also find for 
the higher dimensions investigated.
Here we see that $\gamma(N)$ converges rapidly to a value close to 1.658.
Unfortunately the signal to noise ratio
decreases with $L$ and with dimension; furthermore
the accessible range of $L$ decreases fast as $d$ increases; because of
this, we are able to extract $\gamma$ reliably
only for dimensions 
up to $d=5$. Our results
are summarized in table~\ref{tab:numerics} where the error estimates
come from both statistical noise and
uncertainties in the large $L$ extrapolation.
\begin{table}
    \begin{center}
      \begin{tabular}{ccccccc}
        $~$ & $d=1$ & $d=2$ & $d=3$ & $d=4$ & $d=5$\\
\\
        $\gamma_{hypercube}$  &  1.57082(6)    & 1.6577(6) & 1.7152(10)
    &  1.761(5)  &    1.806(10) \\
        $\gamma_{Cayley}$  &  1.570796  & 1.854075  &  1.927621  & 1.956922
        & 1.971464
      \end{tabular}
    \end{center}
  \caption{\label{tab:numerics} The numerical estimates of $\gamma$ for 
    $d$-dimensional hypercubic lattices ($\gamma_{hypercube}$); error estimates come
    from statistical fluctuations and uncertainties in the large lattice size
    extrapolation. Bottom line: exact values
    for the Cayley tree with the same coordination number, $\mu+1=2d$.}
\end{table}
%


\section{Discussion and Conclusions}
\label{sec:Conclusions}

We have been concerned with the statistical properties
of $M$, the number of local minima in a random energy landscape.
The low order moments of $M$ can easily be obtained;
higher order moments could be computed
by automated counting of graphs. The \emph{atypical} values
of $M$ follow a large deviation principle as given in
Eq.~(\ref{ld1}); we have been able to compute this
function $\Phi(y)$ in one dimension analytically; it diverges
as $y\to 0$ and goes to a finite limit when $M$ 
reaches its maximum value where one half of the 
lattice sites are local minima. This then led us to
consider the limit of maximum packing ($M$ takes
on its largest possible value)
for more general lattices. We focused on bipartite lattices
where up to half of the sites can be local minima.
We derived analytically the probability of this maximum packing
$P(M_{\rm{max}},N)$ for
the Cayley tree and for a two-leg ladder. We then tackled
$d$-dimensional hypercubic lattices by computational techniques.
For all these lattices, it is easy to see that 
$P(M_{\rm{max}},N) \ge 2^{-N}$ simply by forcing the $x_i$ on
the even (odd) sites to be less (greater) than $1/2$;
this immediately leads
to $\gamma(d) \le 2$ for all dimension. Furthermore when $d$ becomes
large, it will be very rare to have maximum packing if
the $x_i$ do not very nearly satisfy this even-odd pattern so one
expects $\gamma(d)$ to tend towards 2 in the large $d$ limit. 

\begin{figure}[htbp]
\includegraphics[width=10cm]{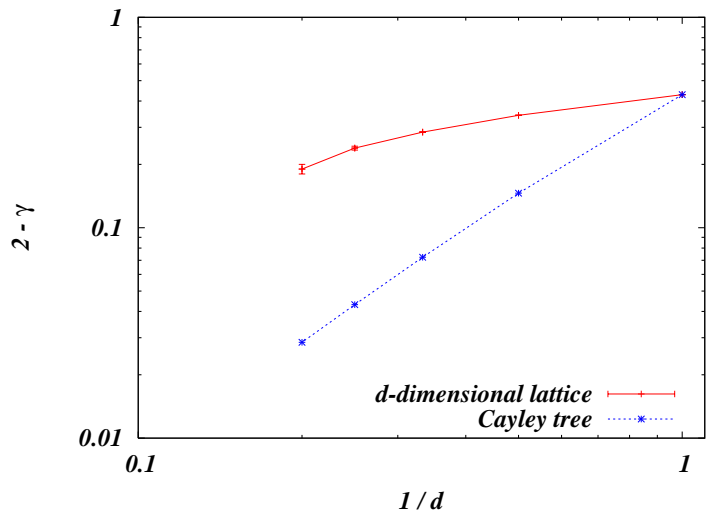}
\caption{The convergence to the large $d$ limit of the 
estimates of $\gamma(d)$ vs $1/d$: plotted is the difference
$2 - \gamma(d)$ which vanishes as a power at large $d$.}
\label{fig:gamma}
\end{figure}

Given this large $d$ limit, It is natural
to ask how $\gamma=2$ is approached. In Fig.~(\ref{fig:gamma})
we show that the Cayley tree case follows a very clear
power correction law which can be derived analytically
as being $1/d^2$. The case of the $d$-dimensional lattice
is less clear but is compatible with a $1/d$ law.

The Cayley tree thus provides an
approximation to the hypercubic case but not a very
accurate one; this is probably because the nature of the correlations
from site to site of the $x_i$ (given that one has a maximally
packed configuration) are quite different when considering
the tree rather than the hypercube.

Our work can be extended in several ways. (i) We computed the exact large 
deviation function in one dimension, but this function can also be 
determined for the Cayley tree~\cite{SollichMajumdar07}. (ii) One 
can also introduce
a chemical potential $z$ for each minimum and consider the
thermodynamics of minima as a function of $z$; recent 
numerical results by Derrida in $2$-d~\cite{Derrida07}
indicate the presence of a phase transition with $z$.
(iii) Our variables $x_i$ were i.i.d. random variables; 
does a large deviation principle still hold if these
variables have short range correlations? One expects so.
(iv) How do all these statistical properties generalize
if one asks for minima within a given energy range?
Finally, it would also be of interest to understand the similarities
and differences between the statistics of local minima
in continuous random energy 
landscapes~\cite{CavagnaGarrahan99,Fyodorov04,Fyodorov05}
and in lattice models as presented here; the energies could
be random or correlated as in the Sinai problem.

\acknowledgments
This work was supported by the EEC's FP6 Information Society
Technologies Programme
under contract IST-001935, EVERGROW (www.evergrow.org), and by the
EEC's HPP under contract HPRN-CT-2002-00307 (DYGLAGEMEM). 
S.M. wants to thank the hospitality of the
Isaac Newton Institute, Cambridge (UK) where this work was completed.
We thank our colleagues A.J. Bray, B. Derrida, D. Dhar, 
M.R. Evans, K. Mallick, S. Nechaev, G. Oshanin and P. Sollich 
for stimulating discussions.

\end{document}